\documentclass[11pt]{article}
\pdfoutput = 1
\usepackage[utf8]{inputenc}

\usepackage[top=1in,bottom=1in,left=1in,right=1in]{geometry}
\makeatletter \g@addto@macro\@floatboxreset\centering \makeatother
\usepackage{color}
\definecolor{darkgreen}{rgb}{0,0.5,0}
\definecolor{darkblue}{rgb}{0,0,0.6}
\definecolor{purple}{rgb}{0.4,.2,0.7}
\usepackage[noEucal]{main}
\usepackage{aas_macros,empheq,fancybox,graphicx,multirow}
\usepackage{aas_macros}
\usepackage{ulem}
\usepackage{esint}
\usepackage{appendix}
\usepackage{mathabx}
\usepackage{tensor} 
\usepackage{footmisc}
\newcommand{\SAN}[1]{{\color{purple}[SN: #1]}}

\newcommand{\nn}{\nonumber}

\begin{document}

\thispagestyle{empty}

\begin{center}
    ~
    \vskip10mm

     {\LARGE  {\textsc{Soft Scalars and the Geometry of the Space of Celestial CFTs}}}
    \vskip10mm
    
Daniel Kapec,$^{a,b}$  Y.T. Albert Law,$^b$ and  Sruthi A. Narayanan$^{b}$ \\
    \vskip1em
    {\it
        $^a$ Center of Mathematical Sciences and Applications, Cambridge, Massachusetts 02138, USA\\ \vskip1mm
        $^b$ Center for the Fundamental Laws of Nature,
Harvard University, Cambridge, Massachusetts 02138, USA\\ \vskip1mm
         \vskip1mm
    }
    \vskip5mm
    \tt{danielkapec@fas.harvard.edu, ylaw1@g.harvard.edu, sruthi$\_$narayanan@g.harvard.edu}
\end{center}
\vspace{10mm}

\begin{abstract}
\noindent
Known examples of the holographic dictionary in asymptotically Anti-de Sitter spacetimes equate moduli spaces of bulk vacua with conformal manifolds in the dual quantum field theory. We demonstrate that the same identification holds for gravity in asymptotically flat spacetimes in any dimension, in accord with expectations derived from the celestial conformal field theory (CCFT) formalism. Soft limits of moduli scalars described by the sigma model are universal, and relate to parallel transport of $S$-matrix observables over the moduli space of bulk vacua. The leading ``soft moduli operator'' is the shadow transform of a dimension $\Delta=d$ marginal operator $M(x)$. The universal form of the soft limit guarantees that $M(x)$ acts as a marginal deformation in the CCFT$_d$, and coherent states of the soft scalars correspond to finite deformations along the conformal manifold. This manifold typically has curvature, which is captured by the antisymmetric double-soft theorem and which reflects the Berry curvature in CCFT$_d$. We also compute the Mellin-transformed four-point function in the sigma model and compare to a formula of Kutasov for the curvature of the conformal manifold.

\end{abstract}
\pagebreak


\setcounter{tocdepth}{2}
{\hypersetup{linkcolor=black}
\small
\tableofcontents
}

\section{Introduction}
The ``celestial holography'' program  seeks to construct a $d$-dimensional Euclidean ``celestial'' conformal field theory (CCFT) capable of defining (non-perturbatively) the $(d+2)$-dimensional quantum gravity scattering amplitudes in asymptotically flat spacetimes. However, there is as yet no independent construction or definition of such a theory, and the general set of axioms satisfied by CCFTs is not yet known. Thankfully, even in the absence of a concrete realization of such a model, it is still possible to explore the universal, model independent properties shared by all CCFTs.

Conformal field theories have a small set of universal operators. For instance, every local CFT comes equipped with a stress tensor $T_{ab}(x)$ that generates conformal transformations. The existence of this operator in CCFT is guaranteed  by the subleading soft graviton theorem, which explicitly provides an operator satisfying all of the local Ward identities of a stress tensor in CFT$_d$~\cite{Kapec:2017gsg}. Similarly, when the Hilbert space of a CFT exhibits global symmetries, the local operator spectrum includes a conserved current $J_a(x)$. This family of operators is furnished by the leading soft photon, soft gluon, and soft graviton theorems~\cite{Kapec:2017gsg,Kapec:2021eug,Guevara:2021abz}. The existence of these operators is required in order for the global symmetries of the bulk theory and its putative dual to agree.

Beyond this set of universal local operators, much remains to be understood about the operator content and abstract characterization of CCFT. One common method used to investigate an intractable system is to study the model's space of possible deformations. In the context of CCFT,
global conformal symmetry corresponds to Lorentz invariance in the bulk and consequently cannot be violated: the only interesting deformations are those that preserve conformal invariance. We are thus led to study the space of exactly marginal deformations, or equivalently the conformal manifold, of CCFT. 

In standard examples of the holographic correspondence, marginal deformations in the dual CFT map onto continuous moduli spaces of vacua in the bulk gravitational theory. Families of inequivalent bulk vacua are relatively commonplace in models of quantum gravity in asymptotically flat spacetimes, where they appear as K\"{a}hler and complex moduli spaces of supersymmetric string compactifications. This provides a rich set of concrete, self-consistent examples to which our analysis will apply. Indeed, it may turn out that conformal manifolds are more common in celestial CFT than they are in the more conventional CFT duals to quantum gravity in AdS space, where moduli spaces of inequivalent vacua are relatively less common. 

In asymptotically flat spacetimes, the vacuum is determined by boundary conditions (vacuum expectation values, or vevs) at spatial infinity. Long-wavelength fluctuations about these vevs are described by a sigma model with target space given by the vacuum manifold $\mathcal{M}$. The space $\mathcal{M}$ comes equipped with an intrinsic geometry that is specified by the curvature of the metric on the moduli space of vacua. As we will see, this corresponds precisely to the geometry of the space of CCFTs. Infinitesimal variations along the bulk moduli space are captured by long wavelength (soft) scalars, whose $S$-matrix elements are universal and controlled by the moduli space geometry. These soft scattering states define distinguished operators in CCFT, whose role is to generate marginal deformations along the conformal manifold. 

Our analysis provides the first robust structural result in CCFT that does not follow from symmetries of the problem, and enlarges the set of known universal operators $(T_{ab}(x)$ and $J_a(x))$ to include the spin zero sector. The analysis is possible because the geometric soft theorems~\cite{Cheung:2021yog}, although not fixed by symmetry, are totally determined by the geometry of the vacuum manifold. The dual description is simple because the vacuum manifold is defined at spatial infinity and is explicitly a boundary quantity. 

Recently, a number of works considered celestial amplitudes in non-trivial backgrounds that break translation invariance while preserving Lorentz invariance \cite{Fan:2022vbz, Casali:2022fro}. These constructions can be viewed as generalized marginal deformations of CCFT which do not preserve the full set of symmetries that distinguish CCFT from ordinary CFT. It would be interesting to understand the conformal manifolds associated to this more general set of deformations.

The organization of this paper is as follows. In section \ref{sec:Kinematics} we review the celestial CFT formalism. Section \ref{sec:sigmamodel} collects the relevant facts about sigma models and geometric soft theorems.  Section \ref{sec:PerturbationTheory} reviews conformal perturbation theory and establishes its relationship with the geometric soft theorems. In section \ref{sec:Correspondence} we compute the Mellin transform of the four point function of moduli scalars and compare to well known formulas for the curvature tensor of the conformal manifold.

\section{Kinematics and celestial CFT$_d$}\label{sec:Kinematics}
The Lorentz group in $(d+2)$ dimensions is isomorphic to the Euclidean conformal group in $d$ dimensions. This correspondence allows one to rewrite the $(d+2)$-dimensional $S$-matrix as a collection of $d$-dimensional conformal correlators~\cite{Pasterski:2017kqt,Kapec:2017gsg}, as we now briefly review. 

The Lorentz generators $M_{\mu \nu}$ $(\mu,\nu \in 0, \dots d+1)$ in $d+2$ dimensions are linear combinations of the $d$-dimensional Euclidean conformal group generators 
\begin{equation}
\begin{split}\label{confgen}
J_{ab} = M_{ab} \;, \qquad D = M_{d+1,0}  \;, \qquad T_a = M_{0,a} - M_{d+1,a} \;, \qquad K_a = M_{0,a} + M_{d+1,a}  \; . 
\end{split} 
\end{equation}
The $J_{ab}$ generate $SO(d)$ rotations, $T_a$ and $K_a$ generate translations and special conformal transformations respectively, and $D$ is the dilation operator. In these variables the Lorentz algebra
\begin{equation}
    [ M_{\mu\nu} , M_{\rho\s} ] =  i(  \eta_{\rho\mu}M_{\nu\sigma}-\eta_{\rho\nu}M_{\mu\sigma}+\eta_{\sigma\nu}M_{\mu\rho}-\eta_{\sigma\mu}M_{\nu\rho})
\end{equation}
is simply that of the Euclidean conformal group
\begin{gather}\label{eq:confalgebra}
[  J_{ab} ,  J_{cd} ] =  i(\delta_{ca}J_{bd}+\delta_{db}J_{ac}-\delta_{cb}J_{ad}-\delta_{da}J_{bc})   \;, \nonumber\\
[ J_{ab} ,  T_c ] =  i (\delta_{ca}T_b -\delta_{cb}T_a)  \;, \qquad [ J_{ab} ,  K_c ] =  i(\delta_{ca}K_b -\delta_{cb}K_a)  \;, \nonumber \\
[ T_a , D  ] =  i  T_a \;, \qquad [ K_a , D  ] = - i  K_a   \;, \qquad [ T_a ,  K_b ] = - 2 i ( \d_{ab}  D +  J_{ab} ) \;. 
\end{gather}
These generators act on momentum space and preserve the null cone. The action is particularly simple if we parameterize the on-shell momentum as\footnote{There is a more general parameterization of the null momenta familiar from the embedding space formalism: $p^\mu(\o,x) = \o \,\Omega(x)\, {\hat q}^\mu ( \o , x )$. Here $\Omega(x)^2$ is the conformal factor of the conformally flat cross section of the null cone in $d+2$ dimensions \cite{Kapec:2017gsg}. When $d=2$, most references choose coordinates $x_1+ix_2=z$ and $\Omega=2(1+z\bar{z})^{-1}$, corresponding to a spherical (rather than flat) cross-section.}
\begin{equation}\label{eq:mompar}
p^\mu(\o,x) = \o {\hat p}^\mu ( \o , x ) \; , \qquad {\hat p}^\mu(\o,x) = {\hat q}^\mu(x) + ( m^2 / \o^2 )  n^\mu \; ,
\end{equation}
where $\hat{q}$ and $n$ are null vectors
\begin{equation}\label{eq:nullMom}
{\hat q}^\mu(x) = \frac{1}{2} \left( 1 + x^2 , 2 x^a , 1 - x^2 \right) \; , \qquad n^\mu = \frac{1}{2} \left( 1 , 0^a , - 1 \right)  \; . 
\end{equation}
Bulk Lorentz transformations are realized as ``boundary'' conformal transformations of the $x^a$ coordinates. The $S$-matrix is defined to be the overlap of in- and out- scattering states, and the $n$-particle scattering amplitude takes the form 
\begin{equation}
\begin{split}\label{eq:amp}
\CA_n &=  \braket{ p_1 , \cdots , p_m }{ p_{m+1} , \cdots , p_n } \\
&= \bra{0} T \{ a_1^\text{out}(p_1) \cdots a_m^\text{out}(p_m) a_{m+1}^{\text{in}\dagger} (p_{m+1} ) \cdots a_{n}^{\text{in}\dagger} (p_n ) \} \ket{0} \; , 
\end{split}
\end{equation}
where $T$ denotes time ordering. If we  define the operators 
\begin{equation}\label{eq:OperatorDef}
\mathcal{O}_i(\omega_i,x_i) \equiv a_i^\text{out}(p(\omega_i,x_i)) \t(\omega_i) +  \bar{a}_i^{\text{in}\dagger}(-p(\omega_i,x_i)) \t(-\omega_i) \;,
\end{equation}
then this amplitude can be written in the suggestive form
\begin{equation}\label{eq:ConformalRep}
\CA_n =  \avg{ \mathcal{O}_1(\omega_1,x_1) \cdots \mathcal{O}_n(\omega_n,x_n) } \; . 
\end{equation}
In this paper we will be concerned primarily with scalar states, so we will not exhibit spin indices unless necessary. Photon and graviton operators are denoted with spin indices by $\mathcal{O}_a$ and $\mathcal{O}_{ab}$, respectively.
The transformation properties of the $\mathcal{O}(\omega,x)$ are determined by Lorentz invariance. For scalars one has
\begin{equation}\label{eq:Transformation}
[ a_i(p) , M_{\mu\nu} ] = \CL_{\mu\nu}  \,  a_i(p) \; ,  \qquad   \CL_{\mu\nu} = -  i \left( p_{\mu} \p_{p^{\nu}}-p_{\nu} \p_{p^{\mu}}\right)\;  .
\end{equation}
Plugging \eqref{eq:mompar} into \eqref{eq:Transformation}, one finds the transformation properties for massless scalar states
\begin{equation}
\begin{split}
\label{eq:masslessLT}
[ \CO_i(\o,x) , T_a ] &=  i \p_a \CO_i(\o,x) \; , \\
[ \CO_i(\o,x) , J_{ab} ] &=  - i ( x_a  \p_b - x_b \p_a ) \CO_i(\o,x) \; , \\
[ \CO_i(\o,x) , D ] &= i ( x^a \p_a - \o \p_\o ) \CO_i(\o,x) \;  , \\
[ \CO_i(\o,x) , K_a ] &= i  [ x^2  \p_a - 2 x_a x^b \p_b + 2 x_a \o \p_\o  ] \, \CO_i(\o,x) \;  ,
\end{split}
\end{equation}
where $\partial_a\equiv \partial_{x^a}$. This is the transformation law for a conformal primary in CFT$_d$, with the important caveat that the operator $D$ is not diagonal in this basis. Instead, the scattering operators have a formal scaling dimension $\Delta=-\omega \partial_\omega$. This simply reflects the fact that momentum eigenstates are not simultaneously boost eigenstates. For massless particles, this can be fixed by performing a Mellin transform
\begin{equation}\label{eq:MellinTransform}
\widehat{\mathcal{O}}^\pm(\Delta,x) = \int_{\mathcal{C}} d\omega \, \omega^{\D-1} \mathcal{O}(\pm\omega,x) \; . 
\end{equation}
The correlation functions of these operators transform as conformal correlators in a Euclidean CFT$_d$ provided that the integrals converge. We have purposefully not specified the contour of integration or the admissible values of $\Delta$ in \eqref{eq:MellinTransform}.
Many works on this subject assume that the correct conformal weights lie on the principal series $\Delta\in\frac{d}{2}+i\mathbb{R}$ and that the contour is noncompact: 
\begin{equation}
\label{eq:}
\widehat{\mathcal{O}}^\pm(\Delta,x) = \int_0^\infty d\omega \, \omega^{\D-1} \mathcal{O}(\pm\omega,x) \; , \qquad \D \in \frac{d}{2} + i \mrr \; . 
\end{equation}
This choice has an obvious shortcoming in that it excludes the integer-valued scaling dimensions of crucial operators like the stress tensor and conserved currents. Indeed, a number of works (particularly those dealing with universal properties of CCFT) have demonstrated that a more general class of operators, including those with integer dimensions, is required \cite{Kapec:2017gsg,Kapec:2021eug,Atanasov:2021oyu,Guevara:2021abz,Himwich:2021dau}. 
This class of operators arises as ``residue operators'' with integer dimensions and compact Mellin contours surrounding the origin \cite{Kapec:2021eug}
\begin{equation}\label{eq:compactcontour}
\widehat{\mathcal{O}}(n,x) =\oint_\CC \frac{d\omega}{2\pi i }  \,\omega^{n-1}\mathcal{O}(\omega,x) \; . 
\end{equation}
These operators are intrinsically related to soft scattering states since they isolate the individual terms in the soft Laurent-expansion of scattering amplitudes. 
For instance, the leading soft-photon operator
\begin{equation}
    S_a(x)= \oint \frac{d\omega}{2\pi i}\mathcal{O}_a(\omega,x) 
\end{equation}
isolates the Weinberg pole in abelian gauge theory, while the subleading soft graviton operator 
\begin{equation}
    S_{ab}(x)=\oint \frac{d\omega}{2\pi i }\omega^{-1}\mathcal{O}_{ab}(\omega,x)
\end{equation}
isolates the subleading, $O(1)$ term in the soft graviton expansion. The role of these operators in CCFT is to furnish conserved currents and the stress tensor via the shadow transform \cite{Kapec:2017gsg,Kapec:2021eug}. The shadow transform maps a conformal primary of dimension $\Delta$ and spin $s$ to an operator of dimension $d-\Delta$ and spin $s$.
For an operator in the representation $\mathcal{R}$ of $SO(d)$ it takes the form
\begin{equation}
    \widetilde{ \CO}(x) \equiv \int d^d y \frac{ 1 }{ [ ( x - y )^2 ]^{d-\D} } \mathcal{R}  ( \CI(x-y) ) \cdot \CO(y) \; ,
\end{equation}
where $\mathcal{I}(x)$ is the conformally covariant tensor.  For scalar operators this is simply
\begin{equation}
\begin{split}\label{eq:shadow}
\widetilde{ \mathcal{O}}(x) \equiv 
\int d^d y \frac{ 1 }{ [ ( x - y )^2 ]^{d-\D} }  \mathcal{O}(y) \; .
\end{split}
\end{equation}
Repeated shadow transforms are proportional to the identity\footnote{Note that this coefficient has a finite limit for $s=0$ and $\Delta \to d$ when $d$ is even, but vanishes linearly in $(\Delta-d)$ when $d$ is odd. Similar statements hold for $s=1$ and $\Delta\to d-1$ as well as for $s=2$ and $\Delta \to d$. In each case, the regulated shadow integral in \eqref{eq:currents} also vanishes, and the combination is finite.} 
\begin{equation}\label{eq:ShadowSquared}
\begin{split}
\widetilde{ \widetilde{ \mathcal{O}}} (x) = c_{\D,s} \mathcal{O}(x)\;  , \qquad c_{\D,s} = \frac{\pi^d (\D-1)(d-\D-1)\G(\frac{d}{2}-\D)\G(\D-\frac{d}{2})}{(\D-1+s)(d-\D-1+s)\G(\D)\G(d-\D)} \; .  
\end{split}
\end{equation}
The shadow transforms of the leading soft photon operator and the subleading soft graviton operator
\begin{equation}\label{eq:currents}
    J_a(x)=\frac{1}{2c_{1,1}} \widetilde{S}_a(x) \; , \qquad T_{ab}(x)=-\frac{1}{c_{0,2}} \widetilde{S}_{ab}(x) \; ,
\end{equation}
define operators in celestial CFT that obey all the Ward identities of a conserved current and stress tensor, respectively \cite{Kapec:2017gsg}. These relationships suggest that bulk scattering operators should have interesting boundary shadow transforms, although the extent of this relationship is not fully understood. We will see that a similar set of statements relates bulk soft scalars to boundary marginal operators, offering support for the general correspondence.

\section{Sigma models and soft theorems}\label{sec:sigmamodel}
In standard examples of the holographic dictionary involving asymptotically AdS spacetimes, moduli spaces of bulk vacua map onto the conformal manifold of the dual CFT. 
Although the dictionary in asymptotically flat space is much less well-understood, we will see in the following sections that a version of this correspondence also holds for flat space holography within the CCFT formalism. 

Continuous spaces of vacua arise frequently in quantum field theory in flat spacetime. In these examples, the moduli space of bulk vacua is parameterized by vacuum expectation values $v^I$, which are in turn determined  by  boundary conditions of local fields at spatial infinity $v^I=\avg{\Phi^I}_{i^0}$. Long-wavelength fluctuations about these vacua are described by a sigma model with target space given by the vacuum manifold $\mathcal{M}$. In more than two dimensions these models are non-renormalizable so this is only an effective description, valid for low energies below some cutoff.

Sigma models arise frequently in superstring compactifications with asymptotically flat factors, where they describe fluctuations of the complex and K\"{a}hler moduli of internal Calabi-Yau manifolds \cite{deBoer:2008ss,Dixon:1989fj}.
Another commonly encountered example of the sigma model arises from spontaneous breaking of a continuous symmetry group $G$ to a subgroup $H$.  In this case, the vacuum manifold is $\mathcal{M}=G/H$ and the scalar excitations about the vacuum are typically termed pions. When $G$ and $H$ arise as global symmetry groups, they are assumed to be compact in order to avoid pathologies.  Although this example is fundamental to quantum field theory in flat space, global symmetries are widely believed to be absent in quantum gravity so the relevance to CCFT is unclear. Vacuum manifolds with $\mathcal{M}=G/H$ and $G$ non-compact also arise frequently as the scalar manifolds in supergravity \cite{Cecotti:2015wqa} and should be relevant for fully self-consistent CCFTs. 

The sigma model is defined as a functional integral over maps into the target manifold
\begin{equation}
    \Phi: \mathbb{R}^{d+1,1}\to \mathcal{M} \; .
\end{equation}
In more than two dimensions $(d>0)$ the model is non-renormalizable, but the minimal action is
\begin{equation}\label{eq:SigmaAction}
    S=\frac12\int d^{d+2}x \; G_{IJ}(\Phi)\partial_\mu \Phi^I\partial^\mu \Phi^J \; .
\end{equation}
The fields $\Phi^I$ are regarded as coordinates on $\mathcal{M}$, and $G_{IJ}(\Phi)$ is the metric on $\mathcal{M}$, which must be Riemannian in order for the kinetic terms to have the correct signs. Field-redefinitions in the quantum field theory correspond to diffeomorphisms of $\mathcal{M}$, and the field-redefinition invariance of physical observables like the $S$-matrix is reflected in the diffeomorphism covariance of the physical quantities calculated from the action \eqref{eq:SigmaAction}.

In order to perform  perturbative calculations in the sigma model, one expands the fields about their vacuum expectation values $v^I$ at spatial infinity, $\Phi^I=v^I + \phi^I$, and path integrates over the normalizable fluctuations.
The action is simplest in Gaussian normal coordinates, and the first few terms are  \cite{Friedan:1980jf}
\begin{align}\label{eq:SigmaActionExpansion}
    S&=\frac12\int \delta_{IJ}\partial\phi^I \partial \phi^J +\frac13 R_{IKLJ}\partial \phi^I \partial \phi^J \phi^K \phi^L
    +\frac16\nabla_KR_{ILMJ}\partial \phi^I \partial \phi^J \phi^K\phi^L\phi^M +\dots
\end{align}
Here $\nabla$ denotes the covariant derivative on $\mathcal{M}$ and $R_{IJKL}$ is the Riemann tensor. In this expansion, $R_{IJKL}$ and its derivatives are functions of the vacuum expectation values $v^I$, which parameterize $\mathcal{M}$.
The tree-level 4-point amplitude of moduli scalars takes the form
\begin{equation}\label{eq:4ptPions}
   A^{I J K L}=s_{34}R^{I K J L} + s_{24}R^{I J K L} \; ,
\end{equation}
where $s_{ij} \equiv (p_i+p_j)^2$.
Higher point amplitudes can also be written in terms of geometric data. For instance, the tree-level five-point amplitude in the sigma model takes the form \cite{Cheung:2021yog}
\begin{align}\label{eq:5ptPions}
    A^{I_1\ldots I_5} = \nabla^{I_3}R^{I_1I_4I_2I_5}s_{45}&+\nabla^{I_4}R^{I_1I_3I_2I_5}s_{35} + \nabla^{I_4}R^{I_1I_2I_3I_5}s_{25} \\ &+ \nabla^{I_5}R^{I_1I_3I_2I_4}s_{34}+\nabla^{I_5}R^{I_1I_2I_3I_4}(s_{24}+s_{45}) \; . \notag
\end{align}

\paragraph{Geometric soft theorem}

There is a long history of soft limits in the sigma model, dating back to the early attempts to re-sum soft-pion emissions in strong-interaction processes \cite{Weinberg:1966kf,Weinberg:1970bs,Weinberg:1972pe}. The results were usefully summarized and extended in the recent work \cite{Cheung:2021yog}, which emphasized the geometric content of the soft theorems. In the notation of section \ref{sec:Kinematics}, the single soft limit of a moduli scalar $\mathcal{O}_I(\omega,x)$ takes the form \cite{Cheung:2021yog}
\begin{equation}\label{eq:singlesoft thm}
    \lim_{\omega\to 0} \;\avg{ \mathcal{O}_I(\omega,x)\mathcal{O}_1(\omega_1,x_1)\dots \mathcal{O}_n(\omega_n,x_n)}_{v} = \nabla_I \; \avg{ \mathcal{O}_1(\omega_1,x_1)\dots \mathcal{O}_n(\omega_n,x_n)}_v \; .
\end{equation}
The subscript $\avg{\cdot}_v$ indicates that the $S$-matrix element is computed with the boundary conditions $\avg{\Phi^I}_{i^0}=v^I$, and the covariant derivative $\nabla$ is understood to act on the $S$-matrix element viewed as a tensorial function of the vacuum expectation values $v^I$ that parameterize $\mathcal{M}$. 

Equation \eqref{eq:singlesoft thm} motivates the definition of the ``leading soft moduli'' operator
\begin{equation}\label{eq:SoftOP}
    S_I(x)\equiv \oint \frac{d\omega}{2\pi i }\omega^{-1}\mathcal{O}_I(\omega,x) \; .
\end{equation}
This operator has conformal dimension $\Delta=0$ and its matrix elements can be calculated from the soft theorem
\begin{equation}\label{eq:singlesoft}
\avg{ S_I(x) \mathcal{O}_1 \cdots \mathcal{O}_n }_v  =  \nabla_I\; \avg{ \mathcal{O}_1 \cdots \mathcal{O}_n }_{v} \; .
\end{equation}
Note that this is essentially a topological operator in the sense that its matrix elements are independent of the position on the celestial sphere. However, multiple insertions of this operator can contain position dependence and we comment on this below.
Since $S_I$ has $\Delta=0$, its shadow transform is formally a marginal operator with $\Delta=d$
\begin{equation}
    M_I(x)=\int d^dy \frac{1}{(x-y)^{2d}}\,S_I(y) \; .
\end{equation}
Inverting this relation using \eqref{eq:ShadowSquared}, it follows that $S_I$ is the shadow transform of a marginal operator
\begin{equation}
    S_I(x) = \lim_{\varepsilon\rightarrow 0}\frac{1}{c_{\varepsilon,0}}\int d^dy\frac{1}{(x-y)^{2\varepsilon}}M_I(y) \; .
\end{equation}
The choice of boundary conditions (the scalar expectation values $v^I$) is part of the definition of the model \eqref{eq:SigmaActionExpansion}, and each choice of vacuum $v^I$ determines a separate Hilbert space and collection of observables that depend explicitly on $v^I$: the Hilbert space and the operators of the model are fibered over the space of vacua $\mathcal{M}$. 
Since the fluctuations $\phi^I$ describe the long-wavelength variations of the vacuum expectation values, the zero-modes of $\phi^I$ simply implement shifts in the $v^I$. In other words, adding a coherent state of zero energy soft pions to a state is equivalent to transporting the model around in the space of vacua. For an infinitesimal deformation this is
\begin{align}\label{eq:transport}
    \avg{ \mathcal{O}_1 \cdots \mathcal{O}_n }_{v-\lambda}
    &=\avg{ \mathcal{O}_1 \cdots \mathcal{O}_n \exp\left[-\lambda^I S_I\right] }_{v}\\
    &\equiv\avg{ \mathcal{O}_1 \cdots \mathcal{O}_n \exp\left[-\lambda^I \int d^dxM_I(x)\right] }_{v}. \notag
\end{align}
At higher orders in the presence of curvature this becomes path dependent and one has to path-order along a specific path $\Gamma$ connecting the two models at the points $v$ and $v'$ in the space of vacua
\begin{align}\label{eq:Pathtransport}
     \avg{ \mathcal{O}_1 \cdots \mathcal{O}_n }^\Gamma_{v'}
    &\equiv\avg{ \mathcal{O}_1 \cdots \mathcal{O}_n \mathcal{P}\exp\left[- \int_\Gamma  S_I d\lambda^I\right] }_{v}. 
\end{align}
Here $\mathcal{P}$ denotes path ordering, which is necessary since the bundle over the moduli space is typically not flat: transport around the space of vacua can have holonomy. Expanding \eqref{eq:transport} to first order, the single soft insertion amounts to an infinitesimal deformation in agreement with \eqref{eq:singlesoft}. 

\paragraph{Consecutive double soft theorem}
Taking the path in \eqref{eq:Pathtransport} to be a loop in the moduli space of vacua computes the holonomy of the parallel transported $S$-matrix, and taking the size of the loop to be infinitesimal isolates the curvature of the connection. This can be viewed as composing two infinitesimal deformations, antisymmetrized in order to close the loop. This effect is captured by the antisymmetric double soft limit, which takes the general form \cite{Cheung:2021yog}
\begin{equation}\label{eq:DoubleSoftComm}
 \left[\lim_{q_I\to 0},\lim_{q_J\to 0} \right]A_{n+2}^{K_1\cdots K_n IJ}=   \left[\nabla^{I}, \nabla^{J} \right]A_n^{K_1\cdots K_n}= \sum_{i=1}^n R\indices{^{IJ} ^{K_i}_K}A_n^{K_1\cdots K\cdots K_n} \; .
\end{equation}
If we view the parallel transport around the moduli space of vacua as an adiabatic variation of the parameters defining the quantum mechanical system \eqref{eq:SigmaAction}, then this formula says that the curvature of the associated Berry connection is simply the curvature of $\mathcal{M}$. 

This antisymmetric double soft limit can be expressed in terms of the residue operators \eqref{eq:SoftOP}
\begin{equation}
 \left[\lim_{q_I\to 0},\lim_{q_J\to 0} \right] \sim    \oint_{\mathcal{C}_1}\frac{d\omega'}{2\pi i\omega'}
        \oint_{\mathcal{C}_2}\frac{d\omega}{2\pi i\omega}\mathcal{O}^I(\omega, x)\mathcal{O}^J(\omega' ,y) \; , 
\end{equation}
where $\mathcal{C}_1$ is a contour surrounding the origin and  the contour $\mathcal{C}_2$ is centered about $\omega'$. The expressions are understood to be $\omega$-ordered: the operator on the contour closest to the origin becomes soft (and acts on the state) first. This is the analog of radial ordering in CFT$_2$, but in the complex $\omega$-plane.

The non-commutativity of the double-soft limit in \eqref{eq:DoubleSoftComm} has a well-known analog in non-abelian gauge theory (and in gravity at subleading order in the soft expansion). The standard claim in the literature is that gauge theory (in the Coulomb phase) and gravity in asymptotically flat space have an infinite dimensional moduli space of vacua corresponding to gauge transformations with non-compact support. Insertions of soft gluons and (subleading) soft gravitons correspond to infinitesimal transport around this infinite dimensional space of vacua.  The metric and curvature on these moduli spaces have not been investigated in detail, but they are certainly not flat and it seems likely that the view taken in this paper applies to these more complicated examples. Noncommutaivity of the multi-soft limits would then be reinterpreted as a reflection of the nontrivial curvature of the moduli space of vacua, and the different choices of soft gluon limits would correspond to path dependence of the parallel transport of $S$-matrix elements about this space.

\paragraph{Higher orders and finite deformations}
Expanding \eqref{eq:Pathtransport} out order by order relates the $N$-ple soft theorem for an $(n+N)$-point amplitude to an order-$N$ deformation for an $n$-point correlator at a fixed point in the moduli space 
\begin{equation}
    \avg{\mathcal{O}_1(x_1)\dots \mathcal{O}_n(x_n)}_{\,v-\lambda}=\sum_{N=0}^{\infty}\frac{(-1)^N}{N!}\avg{ \mathcal{O}_1(x_1)\dots \mathcal{O}_n(x_n)(\lambda^I S_I)^N }_{\,v} \; .
\end{equation}
In this language, the ambiguity in multiple soft limits is fixed by the specification of the path.

\paragraph{Adler zero}

The preceding discussion is general and applies to any vacuum manifold $\mathcal{M}$ and any metric on $\mathcal{M}$. If we place restrictions on $G_{IJ}$ and $\mathcal{M}$, then we can say more. For instance, when the Riemann tensor is covariantly constant (i.e. the metric on $\mathcal{M}$ is locally symmetric), the single soft limit \eqref{eq:singlesoft} actually vanishes. This is known in the literature as the ``Adler zero'' \cite{Adler:1964um}; it was discovered in models of spontaneous symmetry breaking, where the metric on $\mathcal{M}=G/H$ is locally symmetric. We will see in section \ref{sec:PerturbationTheory} that this is an interesting example in celestial CFT that imposes restrictions on the operator content and scaling dimensions of the model.

\section{Conformal perturbation theory and the shadow transform }\label{sec:PerturbationTheory}
In this section we discuss conformal perturbation theory in celestial CFT and demonstrate that the structure of the celestial conformal manifold is directly related to the geometry of the moduli space of bulk vacua described by the sigma model. We then reinterpret the soft theorem results discussed in section \ref{sec:sigmamodel} as a deformation of the celestial CFT by an exactly marginal operator. 

The abstract characterization of a conformal field theory includes
the spectrum of local operators $\mathcal{O}_i(x)$, their conformal dimensions $\Delta_i$, and the collection of OPE coefficients $c_{ijk}$. 
In some cases, the spectrum of local operators includes a subset of marginal operators $M_I(x)$ with spin zero and $\Delta=d$. In this case, the combinations 
\begin{equation} \label{eq:MarginalShadow}
    S_I=\int d^dx\, M_I(x)
\end{equation}
are formally conformally invariant. In the context of celestial CFT, it will be important to note that this formula defining the ``operator'' $S_I$ is a special case of the shadow transform \eqref{eq:shadow}. In fact, $S_I$ can be thought of as the shadow of a marginal operator, independent of any relation to celestial CFT.

If the original CFT has a Lagrangian description, then one can add $S_I$ to the action
\begin{equation}\label{eq:Actiondeformation}
    S(\lambda)=S_0+\lambda^IS_I
\end{equation}
and define a deformed model by functional integration. If the original CFT is non-Lagrangian, then the observables in the deformed theory are simply defined to be
\begin{align}\label{eq:deformation}
    \avg{\mathcal{O}_1(x_1)\dots \mathcal{O}_n(x_n)}_{v-\lambda}&=\avg{ \mathcal{O}_1(x_1)\dots \mathcal{O}_n(x_n)e^{-\lambda^I S_I}}_{\,v}  \\
    &= \avg{ \mathcal{O}_1(x_1)\ldots \mathcal{O}_n(x_n)e^{-\lambda^I\int d^dx M_I(x)}}_{\,v} \; . \notag
\end{align}
In this formula, $\langle \cdots \rangle_{\,v}$ denotes a correlation function in the undeformed theory: it is evaluated using the OPE and spectral data of the original CFT. 

When we do finite deformations in multiple directions and there is curvature in the conformal manifold, this equation acquires path dependence at higher orders and becomes
\begin{align}\label{conformal path}
    \avg{\mathcal{O}_1(x_1)\dots \mathcal{O}_n(x_n)}^\Gamma_{v'}&=\avg{ \mathcal{O}_1(x_1)\dots \mathcal{O}_n(x_n)\mathcal{P}\left[e^{-\int_\Gamma d\lambda^I S_I}\right]}_{\,v}  \; , 
\end{align}
where $\Gamma$ is a path that connects the points $v$ and $v'$ on the conformal manifold.

In either case, since the ``operator'' $S_I$ has $\Delta=0$, this deformation has the potential to preserve conformal invariance. If it does (i.e. the operators $M_I(x)$ are \textit{exactly marginal}), then we say that the $\lambda^I$ are local coordinates on the \textit{conformal manifold} $\mathcal{M}$.
Exact marginality is by no means guaranteed, since the CFT data $\{\Delta_i,c_{ijk} \}$ generically changes under a marginal perturbation. 

Conformal perturbation theory approximates the quantity \eqref{eq:deformation} using the series
\begin{equation}\label{eq:PertSeries}
    \avg{\mathcal{O}_1(x_1)\dots \mathcal{O}_n(x_n)}_{\,v-\lambda}=\sum_{N=0}^{\infty}\frac{(-1)^N}{N!}\avg{ \mathcal{O}_1(x_1)\dots \mathcal{O}_n(x_n)(\lambda^I S_I)^N }_{\,v} \; .
\end{equation}
The convergence of this series is not understood even in standard, well-behaved CFTs, but it is certainly an asymptotic series when the original CFT is free. Obviously the situation is even less clear in celestial CFT. 

\paragraph{First order deformation}

Proceeding formally, to first order in perturbation theory one has
\begin{equation}\label{eq:firstOrder}
  \avg{\mathcal{O}_1(x_1)\dots \mathcal{O}_n(x_n)}_{\,v-\lambda}=\avg{ \mathcal{O}_1(x_1)\dots \mathcal{O}_n(x_n)}_{\,v} -\lambda^I \avg{ \mathcal{O}_1(x_1)\dots \mathcal{O}_n(x_n) S_I }_{\, v} \; + O(\lambda^2) \; .
\end{equation}
Rewriting this formula in the limit of an infinitesimal deformation, one therefore concludes
\begin{equation}\label{eq:FirstOrderNabla}
\begin{split}
    \avg{\mathcal{O}_1(x_1)\dots \mathcal{O}_n(x_n) S_I }_{\, v}&= \lim_{\lambda\to 0} \frac{\avg{ \mathcal{O}_1(x_1)\dots \mathcal{O}_n(x_n)}_{\,v}-\avg{ \mathcal{O}_1(x_1)\dots \mathcal{O}_n(x_n)}_{\,v-\lambda}}{\lambda^I}\\
    &= \nabla_I\, \avg{\mathcal{O}_1(x_1)\dots \mathcal{O}_n(x_n)}_{\,v} \; .
    \end{split}
\end{equation}
This is to be compared with \eqref{eq:singlesoft}. Formulas of this form require careful interpretation. In particular, correlation functions involving one or more insertions of the operator $S_I$, such as \eqref{eq:FirstOrderNabla}, involve an integrated correlation function in the undeformed theory. Since correlation functions are singular at coincident points, the quantities on the right-hand-side of \eqref{eq:PertSeries} are typically infinite and must be regularized. They are therefore scheme dependent, and the scheme essentially amounts to a definition of the correlation functions of the seemingly nonlocal operators $S_I$. 

Interestingly, bulk scattering amplitudes translated into celestial CFT lead to natural, finite correlation functions of the $S_I$ which do not appear to require operator renormalization or subtraction of infinities.\footnote{It may be the case that different prescriptions for simultaneous soft limits are related to scheme dependence in celestial CFT, since they reflect path dependence in the perturbation theory.} From the bulk point of view this is natural since the $S$-matrix is a physical observable free from ambiguity, but the analogous statement in celestial CFT seems nontrivial.

\paragraph{Second order deformation}

The fact that the curvature of the conformal manifold $\mathcal{M}$ does not vanish means that the holonomy (Berry phase) around closed loops in the conformal manifold is non-trivial  \cite{Chaudhuri:1988qb,Ranganathan:1992nb,Ranganathan:1993vj,Kampf:2013vha,Baggio:2014ioa,Baggio:2017aww,Behan:2017mwi}.  
By definition, the correlation function is a tensor on a vector bundle over $\mathcal{M}$. Parallel transport around an infinitesimal closed loop yields the usual formula for the leading nontrivial holonomy in terms of the curvature 
\begin{equation}\label{eq:PertCommutator}
    [\nabla^I,\nabla^J]\, \avg{ \mathcal{O}_1 \cdots \mathcal{O}_n }= R^{IJ} \, \avg{ \mathcal{O}_1 \cdots \mathcal{O}_n} \; ,
\end{equation}
where $R_{IJ}$ is the curvature two-form.  This is to be compared to \eqref{eq:DoubleSoftComm}.

\paragraph{Exact marginality}
Although the combination \eqref{eq:MarginalShadow} is formally conformally invariant, the deformation \eqref{eq:deformation} is not automatically guaranteed to produce conformally invariant correlation functions. This is because the conformal dimensions of operators generically vary as a function of the $\lambda^I$. For instance, the first-order change in the conformal dimension of a primary $\mathcal{O}_j$ due to an infinitesimal deformation by the marginal operator $M_I(x)$ is
\begin{equation}
    \nabla_I \Delta_{j}\sim c_{Ij j} \; .
\end{equation}
In order for higher-order corrections to remain conformally invariant, the marginal operator needs to remain marginal at leading order and should not pick up an anomalous dimension under the deformation. Vanishing of the three-point functions of the marginal operators
\begin{equation}\label{eq:exactMarginal}
    c_{IJK}=0
\end{equation}
 is therefore a necessary condition for exact marginality. 
The complete (all-orders) set of relations defining an exactly marginal operator in CFT$_d$ is not known, and most examples of conformal manifolds rely on non-renormalization theorems (following from supersymmetry) for the dimensions of marginal operators. As we will see, the situation is somewhat better in celestial CFT.

Three-point functions (and two-point functions) are still not completely understood within the celestial CFT formalism. For instance, massless kinematics in $(1,d+1)$ signature sets the three-point on-shell amplitudes of gluons and gravitons to zero, and the three-point amplitude for derivative-coupled scalars also vanishes identically for on-shell momentum-preserving kinematics.\footnote{It has been suggested that to circumvent this, one should study the corresponding quantities in $(2,d)$ signature where the analog of the on-shell three-point amplitude does not vanish \cite{Atanasov:2021oyu,Melton:2021kkz}.} One is tempted to identify this as a reflection of the exact marginality condition \eqref{eq:exactMarginal}, but that is not quite correct. 
Exact marginality is equivalent to the preservation of conformal invariance, and a non-trivial bulk three point amplitude (arising from a scalar potential) is certainly consistent with Lorentz (conformal) invariance. Therefore exact marginality cannot simply be a consequence of the vanishing of the tree-level on-shell three-point function. Rather, the existence of a non-trivial conformal manifold (bulk moduli space of vacua) is equivalent to the fact that the scalar spectrum is non-perturbatively  gapless.  

\paragraph{Intrinsic geometry of the conformal manifold}
When there are exactly marginal operators, the intrinsic geometry of the conformal manifold can be described using the abstract conformal field theory data. The two-point function of marginal operators defines the Zamolodchikov metric on $\mathcal{M}$
\begin{equation}\label{eq:ZamMet}
    G_{IJ}(v)=\avg{M_I(x)M_J(y) }_{v}(x-y)^{2d} \; .
\end{equation}
This metric is always curved, and we denote the curvature tensor $R_{IJKL}$. The geometry of the conformal manifold is encoded in the OPE of the marginal operators \cite{Seiberg:1988pf,Kutasov:1988xb}
\begin{equation}\label{marginal ope}
   M_I(x)M_J(y) \sim \frac{G_{IJ}(v)}{(x-y)^{2d}} + \Gamma_{IJ}^KM_K(y)\delta^{(d)}(x-y)+ \dots
\end{equation}
Here $\Gamma_{IJ}^K$ is the Levi-Civita connection. Since it enters as a local counterterm, this quantity is scheme dependent and will not appear in scheme independent observables, which are independent of the coodinatization of the moduli space. Another way to say this is to note that the integral
\begin{equation}
  \Gamma^K_{IJ} \sim  \frac{\partial g_{IJ}}{\partial \lambda_K}= \int d^dx \; \avg{ M_I(0)M_J(1) M_K(x)} \; 
\end{equation}
is divergent and has to be regulated in a particular scheme. We are free to pick a scheme in which this quantity vanishes identically, which is the analog of choosing the Gaussian normal coordinates in the sigma model \eqref{eq:SigmaActionExpansion}. 

The Riemann tensor can be obtained from a twice integrated four-point function \cite{Kutasov:1988xb,Friedan:2012hi} 
\begin{eqnarray}\label{eq:kutasovformula}
    C_{KLIJ}& = & \int d^dx\, d^dy \;  \avg{ M_K(x)M_{L}(y) M_{I}(1)M_J(0)} \;, \; \cr
    R_{IJKL}& = & \frac12 \left( C_{KILJ} -C_{KJLI} + C_{LJKI}-C_{LIKJ}\right) \; .
\end{eqnarray}
In section \ref{sec:Correspondence}, we will try to interpret this quantity as a correlation function 
\begin{equation}\label{eq:CurvShadow}
    C_{KLIJ}=  \avg{ S_K S_L M_{I}M_J} \; 
\end{equation}
of two marginal operators and two shadow transformed $\Delta=0$ operators in celestial CFT.

\paragraph{When the moduli space is locally symmetric}
Kutasov \cite{Kutasov:1988xb} demonstrated that a sufficient condition in $d=2$ is that the OPE of the marginal operators only contains integer dimension operators. Since this condition is related to a well-studied $S$-matrix and the Adler zero, it would be interesting to check this statement (or find its analog) in celestial CFT.

\section{Mellin transform of the tree-level four point amplitude}\label{sec:Correspondence}
The universal relationship between bulk ``soft moduli scalars'' and marginal operators in celestial CFT is completely encapsulated by the detailed agreement between \eqref{eq:singlesoft}, \eqref{eq:DoubleSoftComm} and \eqref{eq:FirstOrderNabla}, \eqref{eq:PertCommutator}. Corollaries of this identification follow from the detailed form of correlation functions in celestial CFT, and can deviate slightly from their analogues in garden variety CFT, since the structure of celestial CFT and its operator product expansion are relatively nonstandard. In particular, given that two- and three-point functions for massless fields vanish in celestial CFT, it is clear that the formulas \eqref{eq:ZamMet}-\eqref{eq:kutasovformula} must take a slightly different form.

With this in mind, in this section we compute a celestial CFT quantity closely analogous to \eqref{eq:kutasovformula} and find close (though not exact) agreement. The (tree-level) celestial four-point function is related to the tree-level four point amplitude \eqref{eq:4ptPions} through the Mellin transform:
\begin{align}\label{4pt celestial}
    \avg{ \mathcal{O}^I_{\Delta_1}(x_1)\mathcal{O}^J_{\Delta_2}(x_2)\mathcal{O}^K_{\Delta_3}(x_3)\mathcal{O}^L_{\Delta_4}(x_4)} \equiv\left(\prod_{i=1}^4 \int_0^\infty d\omega_i \,  \omega_i^{\Delta_i-1} \right)  A^{IJKL} \, \delta^{(d+2)}\left(\sum_{j=1}^4\epsilon_j \omega_j q_j^\mu\right)\, .
\end{align}
The indices $i,j$ label the external particles, while $\epsilon_i=\pm 1$ indicates whether the particle is outgoing or incoming. All quantities in the integrand are parametrized in terms of the coordinates on the celestial sphere using \eqref{eq:mompar} and \eqref{eq:nullMom}. Rescaling $\omega_i \to \omega_i \, \omega_4$ for $i=1,2,3$, we obtain $\left( \sum_{i=1}^4 \Delta_i \right) -1$ powers of $\omega_4$ from the Mellin integration measure, $-(d+2)$ powers from the delta function, and 2 powers from the amplitude, leading to a factor of the form
\begin{align}\label{eq:deltaftn}
	\int_0^\infty d\omega_4 \, \omega_4^{ \sum_{i=1}^4 \Delta_i -d-1}=2\pi\, \delta\left(\sum_{i=1}^4 \Delta_i-d\right) \, .
\end{align}
Therefore, the sum of the conformal dimensions of operators in the four-point function is constrained to be $d$. Motivated by our discussion of conformal perturbation theory and the results of~\cite{Kutasov:1988xb}, we expect that the Riemann tensor on the conformal manifold should be related to a twice integrated four point function of marginal operators. It seems natural to view this as a correlator of two $\Delta=d$ operators and two  $\Delta=0$ shadow operators\footnote{We are implicitly identifying the $\Delta=0$ ``conformally soft'' operator with the residue operator \eqref{eq:SoftOP}. Single insertions of these operators agree, but it has not been demonstrated that multiple insertions are identical.    } as in \eqref{eq:CurvShadow}. However, both combinations of dimensions are apparently forbidden by the delta function constraint \eqref{eq:deltaftn}, which is a consequence of the extra abelian translational (Poincar\'e) symmetry enjoyed by celestial CFT but not ordinary CFT. 

One possible way to circumvent this obstruction is to compute the non-vanishing four point function with one marginal operator and three $\Delta=0$ operators (so that $\sum \Delta_i=d$), and then to perform a $\Delta=0$ shadow transform on the non-zero correlator so that the result formally\footnote{This procedure indicates that taking the shadow transform does not always commute with taking expectation values in the conformal basis. We hope to understand the interpretation of this calculation better. } resembles a correlation function of two marginal operators and two shadows.  In other words, we \textit{define}
\begin{align}\label{eq:4ptshadow}
    \avg{ \widetilde{\mathcal{O}}^I_{d-\Delta_1}(x_1)&\mathcal{O}^J_{\Delta_2}(x_2)\mathcal{O}^K_{\Delta_3}(x_3)\mathcal{O}^L_{\Delta_4}(x_4)}  \equiv   \int  \frac{d^d x_m}{x_{1m}^{2d-2\Delta_1}} \avg{ \mathcal{O}^I_{\Delta_1}(x_m)\mathcal{O}^J_{\Delta_2}(x_2)\mathcal{O}^K_{\Delta_3}(x_3)\mathcal{O}^L_{\Delta_4}(x_4)} \, ,
\end{align}
with one (not-to-be shadowed) operator of dimension $\Delta=d$ and three operators of dimension $\Delta=0$. It turns out that different choices of operator dimensions can lead to finite or divergent results depending on the specific kinematic configurations. We perform a general calculation of \eqref{eq:4ptshadow} with arbitrary $\Delta_i$ in Appendix \ref{shadow}. Here we summarize the results relevant to \eqref{eq:kutasovformula}. 

To be concrete, consider the case $\epsilon_1=\epsilon_2=-\epsilon_3=-\epsilon_4=-1$, where particles 1 and 2 are incoming  while particles 3 and 4 are outgoing. In this case, the correlators 
\begin{align}\label{As12}
    \mathcal{A}^{IJKL}_{12}(x_i) \equiv& \avg{ \widetilde{\mathcal{O}}^I_{d}(x_1)\mathcal{O}^J_{d}(x_2)\mathcal{O}^K_{0}(x_3)\mathcal{O}^L_{0}(x_4)} = \avg{ \mathcal{O}^I_{d}(x_1)\widetilde{\mathcal{O}}^J_{d}(x_2)\mathcal{O}^K_{0}(x_3)\mathcal{O}^L_{0}(x_4)} \nonumber\\
    =&-  \frac{4\pi \delta\left(0\right)}{|x_{12}|^{2d}} u^{d}\int_1^\infty \frac{dW}{W} \frac{(W-1)^{d-1}\left(  R^{IKJL} W- R^{IJKL} \right) }{\left( W^2-(1+u-v)W+u\right) ^{d}}
\end{align}
and 
\begin{align}\label{As34}
    \mathcal{A}^{IJKL}_{34}(x_i) \equiv &\avg{ \mathcal{O}^I_{0}(x_1)\mathcal{O}^J_{0}(x_2)\widetilde{\mathcal{O}}^K_{d}(x_3)\mathcal{O}^L_{d}(x_4)} = \avg{ \mathcal{O}^I_{0}(x_1)\mathcal{O}^J_{0}(x_2)\mathcal{O}^K_{d}(x_3)\widetilde{\mathcal{O}}^L_{d}(x_4)} \nonumber\\
    =& -\frac{ 4\pi \delta\left(0\right)}{|x_{34}|^{2d}} u^{d}\int_1^\infty \frac{dW}{W} \frac{(W-1)^{d-1}\left(  R^{IKJL} W- R^{IJKL} \right) }{\left( W^2-(1+u-v)W+u\right) ^{d}}
\end{align}
are finite (except for the factor $\delta(0)$ due to the delta function \eqref{eq:deltaftn}). The notation $\mathcal{A}^{IJKL}_{ij}$ indicates that the legs $i,j$ are taken to be marginal and we have expressed the correlators in terms of the $d$-dimensional conformal cross ratios $ u= \frac{x_{12}^2x_{34}^2}{x_{13}^2x_{24}^2}$ and $v=\frac{x_{14}^2x_{23}^2}{x_{13}^2x_{24}^2}$. 

We show in Appendix \ref{shadow} that \eqref{As12} and \eqref{As34} are the only choices of marginal operators for which the correlator is finite. When one marginal operator appears in the in-state and the other appears in the out-state, the result is divergent. For example, the integrals
\begin{align}
    \mathcal{A}^{IJKL}_{13} (x_i) = -  \frac{4\pi \delta\left(0\right)}{|x_{13}|^{2d}} \int_1^\infty dW \frac{W^{d-1}(W-1)^{d-1}\left(  R^{IKJL} W- R^{IJKL} \right)}{\left( W^2-(1+u-v)W+u\right) ^{d}}
\end{align}
and 
\begin{align}
    \mathcal{A}^{IJKL}_{14} (x_i) =- \frac{4\pi \delta\left(0\right)}{|x_{14}|^{2d}} v^{d}\int_1^\infty \frac{dW}{W-1} \frac{W^{d-1}\left(  R^{IKJL} W- R^{IJKL} \right)}{\left( W^2-(1+u-v)W+u\right) ^{d}}
\end{align}
diverge logarithmically in the regions $W\to \infty$ and $W\to 1$ respectively.

The four point functions \eqref{As12} and \eqref{As34} do not exhibit the correct algebraic structure of a curvature tensor and therefore cannot be identified with the Riemann tensor of the conformal manifold. However, the combination 
\begin{equation}\label{curvature formula}
\mathcal{R}^{IJKL} = \frac{1}{2}\left(\mathcal{A}^{KILJ}-\mathcal{A}^{KJLI}+\mathcal{A}^{LJKI}-\mathcal{A}^{LIKJ}\right) 
\end{equation}
taken in \cite{Kutasov:1988xb} is indeed proportional to the Riemann tensor $R^{IJKL}$. For example, 
\begin{align}
    \mathcal{R}^{IJKL}(x_i) = R^{IJKL} \left[-  \frac{4\pi \delta\left(0\right)}{|x_{12}|^{2d}} u^{d} \int_1^\infty \frac{dW}{W} \frac{(W-1)^{d-1}\left(  2W- 1 \right) }{\left( W^2-(1+u-v)W+u\right) ^{d}}\right] \; .
\end{align}
This follows from the analogous statement for the momentum-space amplitude \eqref{eq:4ptPions}: 
\begin{align}
    \frac{1}{2}\left(A^{KILJ}-A^{KJLI}+A^{LJKI}-A^{LIKJ}\right) = R^{IJKL} \left(2s_{34}+s_{24} \right) \; .
\end{align}

\paragraph{Discussion} 

Equation \eqref{eq:kutasovformula} is subtle in standard Euclidean CFT: operators at non-coincident points commute so it is naively impossible to obtain a tensor antisymmetric under index exchange. Of course, the twice-integrated four-point function is formally divergent, and it is a scheme-independent combination of the regulated correlators (which do not satisfy the axioms of local CFT) which yields the Riemann tensor. Although the calculation is not identical, we have seen that there is a version of a ``twice integrated four-point function of marginal operators'' in celestial CFT which is also proportional to the Riemann tensor of the conformal manifold. The correlator retains position dependence in contrast to Kutasov's result
\cite{Kutasov:1988xb, Friedan:2012hi}, but the expression is finite and free from ambiguity.  It seems interesting that the bulk $S$-matrix provides a set of regulated, finite answers for conformal perturbation theory in celestial CFT. 
Indeed, the results obtained in this section may indicate an underlying principle in the CCFT formalism. The relation between kinematic configurations and finiteness of the Mellin transformed celestial correlator is reminiscent of the conclusions in~\cite{Fan:2021isc, Crawley:2021ivb}, where it was necessary to define the Hilbert space of incoming (outgoing) states in terms of unshadowed (shadowed) operators.

\paragraph{Higher-point correlators and covariant derivatives of the Riemann tensor}

In standard conformal perturbation theory, the covariant derivative of the Riemann tensor is related to the five-point function of marginal operators integrated three times~\cite{Kutasov:1988xb}. The corresponding quantity in celestial CFT would appear to be a five point correlator of two $\Delta=d$ and three $\Delta=0$ operators. Unsurprisingly, the bulk tree-level 5-point amplitude \eqref{eq:5ptPions} contains the covariant derivative of the Riemann tensor of the vacuum manifold. This suggests a generalization of the above calculation for the case of a five-point correlator. The operator dimensions work out similarly: rescaling $\omega_i \to \omega_i \, \omega_5$ for $i=1,2,3,4$, we have $\left( \sum_{i=1}^5 \Delta_i \right) -1$ powers of $\omega_5$ from the Mellin integration measure, $-(d+2)$ powers from the delta function and 2 powers from the amplitude \eqref{eq:5ptPions}. Performing the $\omega_5$ integral then leads to a factor $\delta\left(\sum_{i=1}^5\Delta_i - d\right)$, so we can choose $\Delta_1=d,\Delta_2=\cdots =\Delta_5=0$ and shadow one of the dimension 0 operators to obtain an analog of Kutasov's formula. An appropriate combination of these correlators will be proportional to the covariant derivative of the Riemann tensor.   

In general, a correlation function of $n$ marginal operators integrated $(n-2)$ times can be thought of as a correlator of two $\Delta=d$ operators and $(n-2)$ $\Delta=0$ operators in celestial CFT. Conformal perturbation theory predicts that this quantity should be related to the $(n-4)$-th derivative of the Riemann curvature, and the action~\eqref{eq:SigmaActionExpansion} makes it clear that the bulk $n$-point amplitude will produce terms with this structure. The Mellin transform of the bulk $n$-point amplitude always includes a factor of  $\delta\left(\sum_{i=1}^n\Delta_i - d\right)$, so we can always pick $\Delta_1=d,\Delta_2=\cdots =\Delta_n=0$ and shadow one of the dimension 0 operators to obtain an analogous formula in celestial CFT.

\section*{Acknowledgements}
We are grateful to Scott Collier, Prahar Mitra, Monica Pate, Ana Raclariu and Andy Strominger for comments on the draft. This work is supported by the Harvard Center of Mathematical Sciences and Applications as well as DOE grant de-sc/0007870. AL  was supported in part by the Croucher Foundation. 

\appendix

\section{Shadowing the four point correlators}\label{shadow}

In this appendix we compute the shadow correlator \eqref{eq:4ptshadow}:
\begin{align}\label{app:4 pt shadow}
    \avg{ \widetilde{\mathcal{O}}^I_{\bar\Delta_1}(x_1)&\mathcal{O}^J_{\Delta_2}(x_2)\mathcal{O}^K_{\Delta_3}(x_3)\mathcal{O}^L_{\Delta_4}(x_4)}  \equiv   \int  \frac{d^d x_m}{x_{1m}^{2\bar\Delta_1}} \avg{ \mathcal{O}^I_{\Delta_1}(x_m)\mathcal{O}^J_{\Delta_2}(x_2)\mathcal{O}^K_{\Delta_3}(x_3)\mathcal{O}^L_{\Delta_4}(x_4)} 
\end{align}
with $\bar \Delta_i \equiv d-\Delta$, and
\begin{align}\label{app:4pt celestial}
    \avg{ \mathcal{O}^I_{\Delta_1}(x_1)\mathcal{O}^J_{\Delta_2}(x_2)\mathcal{O}^K_{\Delta_3}(x_3)\mathcal{O}^L_{\Delta_4}(x_4)} \equiv\left(\prod_{i=1}^4 \int_0^\infty d\omega_i \,  \omega_i^{\Delta_i-1} \right)  A^{IJKL} \, \delta^{(d+2)}\left(\sum_{i=1}^4\epsilon_i \omega_i q_i^\mu\right)\, .
\end{align}
The tree-level four point amplitude \eqref{eq:4ptPions} in terms of coordinates on the celestial sphere reads
\begin{eqnarray}\label{4 pt}
A^{IJKL}(\omega_i,x_i)
  = -\left(R^{IKJL}\epsilon_3\epsilon_4\omega_3\omega_4 x_{34}^2 + R^{IJKL}\epsilon_2\epsilon_4\omega_2\omega_4 x_{24}^2\right)\, ,
\end{eqnarray}
where $R^{IJKL}$ is the Riemann tensor of the target space. The momentum conserving delta function in \eqref{app:4pt celestial} in these coordinates is
\begin{align}\label{d+2 del}
	\delta^{(d+2)}\left(\sum_{i=1}^4\epsilon_i\omega_i q_i^\mu\right) =2\, \delta\left(\sum_{i=1}^4\epsilon_i \omega_i \right) \delta\left(\sum_{i=1}^4\epsilon_i \omega_i  x_i^2 \right) \delta^{(d)}\left(\sum_{i=1}^4\epsilon_i \omega_i  x^a_i \right)\; .
\end{align}
The $d$-dimensional delta function in \eqref{d+2 del} can be used to do the $d$ shadow integrals in \eqref{app:4 pt shadow}. The remaining two delta functions in \eqref{d+2 del} then localize the $\omega_2$- and $\omega_3$-integrals in \eqref{app:4pt celestial}. Rescaling $\omega_1 \to \omega_1 \omega_4$, the $\omega_4$-integral leads to the delta function \eqref{eq:deltaftn}. The result takes the form 
\begin{align}\label{app:shadow inter}
	&\avg{ \widetilde{\mathcal{O}}^I_{\bar\Delta_1}(x_1)\mathcal{O}^J_{\Delta_2}(x_2)\mathcal{O}^K_{\Delta_3}(x_3)\mathcal{O}^L_{\Delta_4}(x_4)} \nonumber\\
	=&  4\pi \,F(\bar \Delta_1,\Delta_2,\Delta_3,\Delta_4,x_i) \,  u^{\frac{2\bar\Delta_1+2\Delta_2-\Delta_3-\Delta_4}{6}}v^{\frac{2\bar\Delta_1-\Delta_2-\Delta_3+2\Delta_4}{6}}\delta\left(\left( \sum_{i=1}^4 \Delta_i \right)-d\right) \, I_1(u,v) \; ,
\end{align}
where
\begin{align}
	F(\Delta_1,\Delta_2,\Delta_3,\Delta_4,x_i) = \prod_{i<j} |x_{ij}|^{\frac{\mathbf \Delta}{3}-\Delta_i -\Delta_j} \, , \qquad \mathbf \Delta = \sum_{i=1}^4 \Delta_i
\end{align}
is a (non-unique) kinematic factor that accounts for the transformation properties of the correlator under conformal transformations. The other factors in \eqref{app:shadow inter} are expressed in terms of the $d$-dimensional conformal cross ratios $ u= \frac{x_{12}^2x_{34}^2}{x_{13}^2x_{24}^2}$ and $v=\frac{x_{14}^2x_{23}^2}{x_{13}^2x_{24}^2}$. Here, $I_1(u,v)$ is the conformally invariant integral 
\begin{align}
    I_1(u,v) \equiv & \int_0^\infty d\omega_1  \, \theta\left( -\epsilon_2\epsilon_4 -\epsilon_1 \epsilon_2\omega_1 \right) \theta\left( -\epsilon_1\epsilon_3 -\epsilon_3\epsilon_4 \omega_1 \right) \left( -\frac{\epsilon_4}{\epsilon_2} -\frac{\epsilon_1}{\epsilon_2} \omega_1 \right)^{\Delta_2-1}\left( -\frac{\epsilon_1}{\epsilon_3} \frac{1}{\omega_1}-\frac{\epsilon_4}{\epsilon_3} \right) ^{\Delta_3-1} \nonumber \\ &\frac{\omega_1^{\Delta_1-d-1}}{\left| Z+\frac{\epsilon_4}{\epsilon_1 \omega_1} \right| ^{2\bar\Delta_1}}\left(  R^{IKJL}\epsilon_3 \epsilon_4  \left( \frac{\epsilon_1}{\epsilon_3} \frac{1}{\omega_1}+\frac{\epsilon_4}{\epsilon_3}\right)+ R^{IJKL}\epsilon_2 \epsilon_4 \left( \frac{\epsilon_4}{\epsilon_2} +\frac{\epsilon_1}{\epsilon_2} \omega_1 \right)  \right) \; ,
\end{align}
where $Z$ is a complex number related to $u,v$ through $u=|Z|^2$ and $v=|Z-1|^2$. Due to the Heaviside step functions in the integrand, we have
\begin{align}
    I_1 (u,v)= 
    \begin{cases}
    I_1^s (u,v) &\quad  \epsilon_1 = \epsilon_2 =- \epsilon_3 =- \epsilon_4 \quad \text{($s$-kinematics)} \\
    I_1^t (u,v) &\quad \epsilon_1 = -\epsilon_2 = \epsilon_3 =- \epsilon_4 \quad \text{($t$-kinematics)}\\
    I_1^u (u,v) &\quad  \epsilon_1 = -\epsilon_2 = -\epsilon_3 =\epsilon_4 \quad \text{($u$-kinematics)}\\
    0 & \quad \text{Otherwise}
    \end{cases}
\end{align}
where
\begin{gather}
	I_1^s = \int_1^\infty dW G_1(W)\, , \qquad I_1^t = \int_0^1 dW G_1(W)\, , \qquad I_1^u = \int_{-\infty}^0 dW G_1(W) \; ,\nn\\
	G_1(W) = \frac{|W|^{\bar\Delta_1-\Delta_2-2}|W-1|^{\bar\Delta_1-\Delta_4}}{\left( W^2-(1+u-v)W+u\right) ^{\bar\Delta_1}}\left(  R^{IKJL} \frac{W^2}{W-1}- R^{IJKL} \frac{W}{W-1}\right) \, .
\end{gather}
This concludes the computation of the integral \eqref{app:4 pt shadow}. To obtain the correlators with the shadow transform on other legs, one simply relabels the expressions at every step above. For example, exchanging $1\leftrightarrow 2$ and $3\leftrightarrow 4$ everywhere,\footnote{Note that the momentum conserving delta function \eqref{d+2 del}, the kinematics and the cross ratios $u,v$ are all invariant under this exchange. The amplitude \eqref{4 pt} is invariant under this exchange on the support of \eqref{d+2 del}.} we obtain the correlator with the second leg shadowed: 
\begin{align}\label{app:shadow inter 2}
	&\avg{ \mathcal{O}^I_{\Delta_1}(x_1)\widetilde{\mathcal{O}}^J_{\bar\Delta_2}(x_2)\mathcal{O}^K_{\Delta_3}(x_3)\mathcal{O}^L_{\Delta_4}(x_4)} \nonumber\\
	=&  4\pi \,F( \Delta_1,\bar\Delta_2,\Delta_3,\Delta_4,x_i) \,  u^{\frac{2\Delta_1+2\bar\Delta_2-\Delta_3-\Delta_4}{6}}v^{\frac{2\bar\Delta_2-\Delta_1-\Delta_4+2\Delta_3}{6}}\delta\left(\left( \sum_{i=1}^4 \Delta_i \right)-d\right) \,  I_2(u,v) \; .
\end{align}
The conformally invariant integral is
\begin{align}
    I_2 (u,v)= 
    \begin{cases}
    I_2^s (u,v) &\quad  \epsilon_1 = \epsilon_2 =- \epsilon_3 =- \epsilon_4 \quad \text{($s$-kinematics)} \\
    I_2^t (u,v) &\quad \epsilon_1 = -\epsilon_2 = \epsilon_3 =- \epsilon_4 \quad \text{($t$-kinematics)}\\
    I_2^u (u,v) &\quad  \epsilon_1 = -\epsilon_2 = -\epsilon_3 =\epsilon_4 \quad \text{($u$-kinematics)}\\
    0 & \quad \text{Otherwise}
    \end{cases}
\end{align}
with
\begin{gather}
	I_2^s = \int_1^\infty dW G_2(W)\, , \qquad I_2^t = \int_0^1 dW G_2(W)\, , \qquad I_2^u = \int_{-\infty}^0 dW G_2(W)\; , \nn\\
	G_2(W) = \frac{|W|^{\bar\Delta_2-\Delta_1-2}|W-1|^{\bar\Delta_2-\Delta_3}}{\left( W^2-(1+u-v)W+u\right) ^{\bar\Delta_2}}\left(  R^{IKJL} \frac{W^2}{W-1}- R^{IJKL} \frac{W}{W-1}\right) \, .
\end{gather}
The correlators with the third or fourth leg shadowed can be obtained in a similar fashion. Because of this, it suffices to consider \eqref{app:4 pt shadow}.

\paragraph{Two marginal and two shadowed marginal}

We now consider the combinations of operator dimensions discussed in the main text. It is clear that we must take $\bar\Delta_1=d$ while we are free to pick one of $\Delta_2,\Delta_3,\Delta_4$ to be $d$ and the other two to be 0. When $\Delta_2=d$ and $\Delta_3=\Delta_4=0$, 
\begin{align}
    G_1(W) = \frac{|W|^{-2}|W-1|^{d}}{\left( W^2-(1+u-v)W+u\right)^{d}}\left(  R^{IKJL} \frac{W^2}{W-1}- R^{IJKL} \frac{W}{W-1}\right) \, .
\end{align}
This expression makes it is clear that $I_1^s$ is finite\footnote{The integral is a sum of Appel $F_1$ functions.} 
while $I_1^t$ and $I_1^u$ diverge logarithmically in the regions $W\to 0$. Similarly, when $\Delta_3=d$ and $\Delta_2=\Delta_4=0$, 
\begin{align}
    G_1(W) = \frac{|W|^{d-2}|W-1|^{d}}{\left( W^2-(1+u-v)W+u\right) ^{d}}\left(  R^{IKJL} \frac{W^2}{W-1}- R^{IJKL} \frac{W}{W-1}\right) \, ,
\end{align}
in which case $I_1^t$ is finite while $I_1^s$ and $I_1^u$ diverge logarithmically in the regions $W\to \pm\infty$ respectively. Finally, when $\Delta_4=d$ and $\Delta_2=\Delta_3=0$, 
\begin{align}
    G_1(W) = \frac{|W|^{d-2}}{\left( W^2-(1+u-v)W+u\right) ^{d}}\left(  R^{IKJL} \frac{W^2}{W-1}- R^{IJKL} \frac{W}{W-1}\right) \, ,
\end{align}
in which case $I_1^u$ is finite while $I_1^s$ and $I_1^t$ diverge logarithmically in the region $W\to 1$. Therefore, in order to obtain a finite result, the two dimension $d$ operators must be both incoming (outgoing) and the two dimension 0 operators must be both outgoing (incoming).

\bibliographystyle{apsrev4-1long}
\bibliography{Bib.bib}
\end{document}